\documentclass[11pt,twoside]{article}

\usepackage{asp2006}
\usepackage{epsf}
\usepackage{graphicx}
\usepackage{lscape}

\usepackage{natbib}
\begin{document}
\title{The Horizontal Magnetic Field of the Quiet Sun:
Numerical Simulations in Comparison to Observations
with Hinode}   %%% Fill in title
\author{O. Steiner, R. Rezaei, and R. Schlichenmaier}   
%%% Fill in author names
\affil{Kiepenheuer-Institut f\"ur Sonnenphysik, 79104 Freiburg, Germany}   
 %%% Fill in author affiliations
\author{W. Schaffenberger}
\affil{Physics and Astronomy Department, Michigan State University,
East Lansing MI 48824, USA}
\author{S. Wedemeyer-B\"ohm}
\affil{Institute of Theoretical Astrophysics, P.O.~Box 1029, 
        Blindern, N-0316, Norway}

\begin{abstract} %%% Abstract to run on from here.
Three-dimensional magnetohydrodynamic simulations of the surface layers of the Sun
intrinsically produce a predominantly horizontal magnetic field in the photosphere. 
This is a robust result in the sense that it arises from simulations with largely different 
initial and boundary conditions for the magnetic field. While the disk-center synthetic 
circular and linear polarization signals agree with measurements from Hinode, their
center-to-limb variation sensitively depends on the height variation of the horizontal and 
the vertical field component and they seem to be at variance with the observed behavior.
\end{abstract}

%%% MAIN BODY OF TEXT GOES HERE. CONSULT "INSTRUCTIONS FOR AUTHORS USING
%%% LATEX2E MARKUP", SECTIONS 2.3-2.6 FOR HELP WITH EQUATIONS, FIGURES,
%%% AND TABLES.

\section{Introduction}   %%% Top level section head (remove "%" symbol)
Observations with the spectropolarimeter of the Solar Optical Telescope 
(SOT) onboard the Hinode space observatory \citep{kosugi+al07} indicate 
that seen with a spatial resolution of 0.3\arcsec, quiet internetwork 
regions harbor a photospheric magnetic field whose mean field strength of its
horizontal component considerably surpasses that  of the vertical component  
\citep{lites+al08,orozco+al07}. According to 
these papers, the vertical fields are concentrated in the intergranular 
lanes, whereas the stronger, horizontal fields  occur most commonly at 
the edges of the bright granules, aside from the vertical fields. 
\cite{harvey+al07} find from recordings with GONG and SOLIS a `seething 
magnetic field' with a line-of-sight component increasing from disk center to 
limb as expected for a nearly horizontal field orientation.
\cite{ishikawa+al08} detected transient horizontal magnetic fields
in plage regions as well. Previously, \cite{martinez_pillet+al97}  and
\cite{meunier+al98} reported observations of weak and strong
horizontal fields in quiet Sun regions. 

Regarding numerical simulations, \citet{ugd+al98}, note
``we find in all simulations also strong horizontal fields above 
convective upflows'', and \citet{schaffenberger+al05,schaffenberger+al06} 
find frequent horizontal fields in their three-dimensional simulations, which 
they describe as ``small-scale canopies''. 
Also the 3-D simulations of \citet{abbett07} display 
``horizontally directed ribbons of magnetic flux that permeate the model 
chromosphere'', not unlike the figures shown by \citet{schaffenberger+al06}.
\citet{schuessler+voegler08} find in a three-dimensional 
surface-dynamo simulation ``a clear dominance of the horizontal field in 
the height range where the spectral lines used for the Hinode observations 
are formed''.  

\cite{steiner+al08} report on results from three-dimensional 
magnetohydrodynamic numerical simulations of the internetwork 
magnetic field with regard to the intrinsically produced horizontal
magnetic field. They compute the polarimetric signal of this field 
and compare it to measurements with Hinode. In the following,
we briefly summarize part of their results and present new calculations 
of the center-to-limb variation of the linear and circular polarization
from these simulations. 

\section{Simulations}
We have carried out three sets of numerical simulations, run v10, run h20,
and run h50, which significantly differ in their initial and boundary conditions
for the magnetic  field. Run v10 starts with a homogeneous vertical 
field of 10~G strength, which is kept vertical at the top and bottom
boundary. These conditions might actually be more appropriate for 
network magnetic  fields because of the preference for one polarity and 
the vertical direction. Run h20 and run h50 start without a magnetic  
field but upwellings that enter the simulation domain across the bottom 
boundary area carry horizontal magnetic  field of a uniform strength of 
20~G and 50~G, respectively, into the box. 
The three-dimensional computational domain extends from 1400 km
below the mean surface of optical depth  $\tau_c = 1$ to 1400 km
above it,  high enough that the top boundary condition should
not unduly tamper the formation layers of the spectral lines that are
used in polarimetric measurements with Hinode.

\begin{figure}[t]
\centering
  \includegraphics[width=0.55\textwidth]{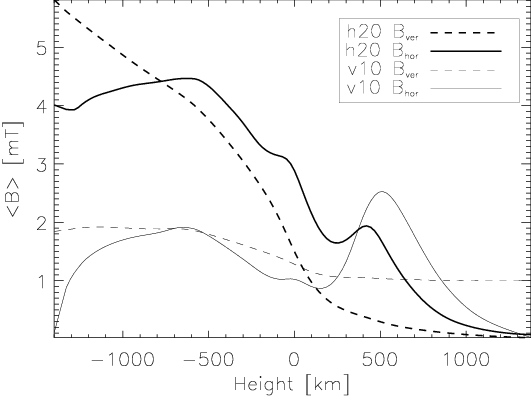}
\caption{Horizontally and temporally averaged horizontal (solid) and 
vertical (dashed) absolute field strength as functions of height for run 
h20 (heavy curves) and for run v10 (light curves).
\label{fig1}}
\end{figure}

Figure~\ref{fig1} shows the horizontally and temporally averaged absolute
vertical and horizontal magnetic field strength as functions of height of
the magnetic field, which emerges from the two runs v10 and h20. In run h20, 
the mean horizontal field strength is larger than 
the mean strength of the vertical component throughout the 
photosphere and the lower chromosphere: in run v10 this is the case in 
the height range between 250~km and 850~km. Both runs show
a local maximum in the horizontal field strength close to the classical 
temperature minimum at a height of around 500~km.
At a given continuum optical depth in the photosphere the area occupied by 
fields with a horizontal component stronger than a given threshold of a few 
gauss is typically three times larger than the area with a vertical component 
exceeding this limit. 

For a comparison with Zeeman measurements from the Hinode spectropolarimeter 
we synthesized the Stokes profiles of both 630~nm \ion{Fe}{i} spectral lines of
the two simulation runs. Profiles were computed with the radiative transfer 
code SIR~\citep{sir92} with  a spectral sampling of 2\,pm. 
When applying an appropriate point spread function, PSF, \citep{wedemeyer08}
to the synthetic profiles and subjecting them to the same procedure for conversion 
to apparent flux densities as done by \cite{lites+al08} for the observed profiles, we
obtain spatial and temporal averages 
for the transversal and longitudinal apparent magnetic flux densities,
$B^{\mathrm{T}}_{\mathrm{app}}$ and $|B^{\mathrm{L}}_{\mathrm{app}}|$  
of respectively 21.5~Mx\,cm$^{-2}$ and 5.0~Mx\,cm$^{-2}$ for run h20 and 
10.4~Mx\,cm$^{-2}$ and 6.6~Mx\,cm$^{-2}$ for run v10.
Thus, the ratio 
$r=\langle  B^{\mathrm{T}}_{\mathrm{app}} \rangle/
   \langle |B^{\mathrm{L}}_{\mathrm{app}}|\rangle = 4.3$ 
for h20 and $1.6$ for v10. \cite{lites+al08} obtain
from Hinode SP data 
$\langle |B^{\mathrm{T}}_{\mathrm{app}}|\rangle = 55$~Mx\,cm$^{-2}$ and 
$\langle |B^{\mathrm{L}}_{\mathrm{app}}|\rangle = 11$~Mx\,cm$^{-2}$ resulting in 
$r = 5.0$.

\section{Center-to-limb variation (CLV)}
\begin{figure}[t]
\centering
  \includegraphics[width=0.48\textwidth]{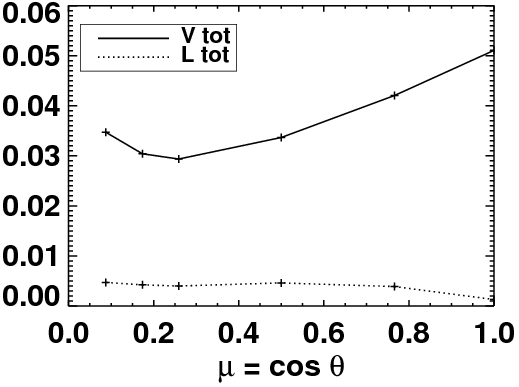}\hfill
  \includegraphics[width=0.48\textwidth]{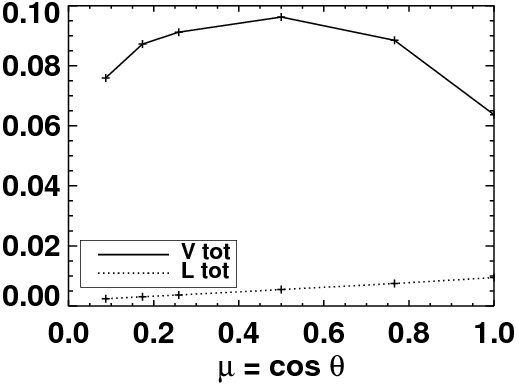}
\caption{Center-to-limb variation (right to left) of the total (wavelength 
integrated) circular ($V_{\mathrm{tot}}$) and linear ($L_{\mathrm{tot}}$) 
polarization in pm of the \ion{Fe}{i}~630.25~nm spectral line of a snapshot 
of run v10 (left) and h50 (right).
\label{fig2}}
\end{figure}

Figure~\ref{fig2} shows the center-to-limb variation of the wavelength integrated
absolute circular ($V_{\mathrm{tot}}$) and absolute linear ($L_{\mathrm{tot}}$) 
polarization signal, 
for the 630.25~nm \ion{Fe}{i} spectral line in pm. The left panel of Fig.~\ref{fig2} 
shows the result for run v10, the right panel for run h50, both for one single 
simulation snapshot. For these results foreshortening was taken into account but no 
PSF. The computational domain of h50 is in the horizontal directions twice as large 
as that for v10 and h20. The CLV of $V_{\mathrm{tot}}$ and 
$L_{\mathrm{tot}}$ substantially differ between v10 and h50, which indicates that 
they sensitively depend on the precise height variation of
the horizontal and the vertical component of the magnetic field. Since the vertical 
component of v10 dominates over the horizontal one in the low photosphere
(cf.~Fig.~\ref{fig1}),
the mean longitudinal (line-of-sight) field component decreases when moving away 
from disk center and only when approaching the limb it increases again due to the 
sharp increase of the horizontal component above $z\approx 200$~km.
This behavior is reflected in the CLV of $V_{\mathrm{tot}}$ of run v10.
The transversal component, viz. the component perpendicular to the line-of-sight
and hence $L_{\mathrm{tot}}$, increases at first because 
the vertical component stays almost constant with height in model v10 (in  
contrast to model h20 or h50). For h50 we have an increase in the longitudinal 
component as a function of disk-center distance because the horizontal component 
dominates throughout the photosphere and as a consequence, the transversal component
decreases. This behavior is reflected by the CLV of $V_{\mathrm{tot}}$ 
and $L_{\mathrm{tot}}$ of run h50.

While the CLV of Fig.~\ref{fig2} finds a natural explanation in terms of the
vertical structure of $\langle B_{\mathrm{hor}}\rangle$ and 
$\langle B_{\mathrm{ver}}\rangle$ shown in Fig.~\ref{fig1}, 
the observations by \cite{lites+al08} tell a different story. Accordingly, the 
circular polarization signal decreases from the disk center to the limb as it 
does in Fig.~\ref{fig2} (left). But also the linear polarization signal decreases 
like in Fig.~\ref{fig2} (right). None of the two simulation snapshots to which
Fig.~\ref{fig2} refers show a simultaneous and monotonic decrease in both quantities.

%\acknowledgements %%% Text of acknowledgements runs on after this command.

%%% THE BIBLIOGRAPHY
%%%
%%% CONSULT SECTION 3 OF "INSTRUCTIONS FOR AUTHORS" FOR HOW TO USE NATBIB.
%%% AUTHORS ARE ENCOURAGED TO USE EITHER THE "THEBIBLIOGRAPY" ENVIRONMENT
%%% BY UNCOMMENTING (DELETING THE "%" SYMBOL) THE COMMANDS BELOW, OR BY
%%% USING THE BIBTEX ENVIRONMENT. TO FIND OUT WHICH IS APPLICABLE TO YOUR
%%% CONTRIBUTION, CONSULT THE VOLUME EDITORS FOR YOUR PROCEEDINGS.
%%%

\end{document}